# Interfaces adaptatives

## Adaptation dynamique à l'utilisateur courant

**Jérôme Simonin et Noëlle Carbonell**

*LORIA, Campus Scientifique*
*BP 239*
*54506 Vandœuvre-lès-Nancy Cedex*
*France*
*{Jerome.Simonin, Noelle.Carbonell}@loria.fr*

RÉSUMÉ. Cet article présente une revue des travaux de recherche récents consacrés à la mise en œuvre en IHM du concept d'adaptation dynamique à l'utilisateur. Nous proposons d'abord une classification des recherches sur les systèmes adaptatifs qui tient compte de la nature des capacités et comportements de l'utilisateur sur lesquels porte la modélisation, de la répartition de l'initiative et du contrôle de l'adaptation entre l'utilisateur et le système, et du rôle de celle-ci. Cette classification est ensuite illustrée par la présentation de quelques travaux de recherche publiés, représentatifs des différentes orientations scientifiques qu'elle distingue. En conclusion, nous évoquons les problèmes que soulève l'étude ergonomique de cette nouvelle forme d'interaction.

ABSTRACT. We present a survey of recent research studies of the implementation of adaptive user models in human-computer interaction. A classification of research directions on adaptive user interfaces is first proposed; it takes account of the user characteristics that are modelled, the distribution of initiative and control of the system evolution between user and system, and the role of dynamic adaptation. Then, a few representative research studies are briefly presented to illustrate this classification. In the conclusion, some major issues regarding the utility and usability of adaptive user interfaces and the design of an appropriate methodology for assessing the ergonomic quality of this new form of interaction are mentioned.

MOTS-CLÉS : Adaptation dynamique à l'utilisateur, Modèles adaptatifs de l'utilisateur, Interfaces adaptatives.

KEYWORDS: Dynamic user adaptation, Adaptive user models, Adaptive user interfaces.

**1. Introduction**

Adapter les interfaces homme-machine logicielles au profil de l'utilisateur courant et au contexte dans lequel se situe son interaction devient une nécessité impérieuse aujourd'hui où le développement de la Société de l'Information accroît considérablement la diversité des utilisateurs et celle des situations d'utilisation. En effet, si l'objectif est de promouvoir une société où l'informatique accompagne dans leurs activités quotidiennes, mobiles ou non, tous les citoyens, quelles que soient leurs capacités cognitives et physiques, leurs connaissances et compétences, leurs motivations et intérêts, il est impératif d'accroître la flexibilité des interfaces utilisateur actuelles, c'est-à-dire leur capacité à évoluer au cours de l'interaction en fonction du profil de l'utilisateur courant et des contraintes qu'induit sur ses capacités perceptives, motrices et/ou cognitives l'environnement dans lequel se déroule l'interaction. La présence de mécanismes d'adaptation du comportement des logiciels interactifs à l'utilisateur courant et au contexte de l'interaction est indispensable, notamment, à la mise en œuvre de concepts tels que 'Accès universel' (*Universal Access*), 'Conception pour tous' (*Universal Design*) d'une part, et 'Informatique mobile/ubiquitaire' (*Pervasive Computing*) ou 'Intelligence ambiante' d'autre part ; pour une définition de ces concepts, voir respectivement (Stephanidis, 2001, a et b), (Weiser, 1996 ; Aarts, Encarnagco, 2006). Globalement, ces mécanismes doivent être capables d'extraire du contexte les informations nécessaires pour créer et mettre à jour un modèle explicite ou implicite de l'utilisateur courant et de son environnement et, plus généralement, de doter le système d'une certaine sensibilité au contexte ou *Context Awareness* ; voir (Shafer et al., 2001 ; Smailagic et al. 2001) pour une définition de ce concept. De la qualité de cette capacité à percevoir et interpréter les informations contextuelles disponibles dans l'environnement, qui constitue l'une des composantes principales de l'intelligence humaine, dépend la pertinence des modèles élaborés et donc la qualité de l'adaptation dynamique de l'interface.

La conception de systèmes interactifs doués de capacités d'adaptation, en particulier d'auto-adaptation, est un domaine de recherche en plein essor qui fait appel à des compétences pluridisciplinaires variées : psychologie, intelligence artificielle et génie logiciel principalement. La difficulté majeure sur laquelle se centrent les efforts actuels est la construction de modèles pertinents et efficaces, qui permettent une adaptation dynamique du comportement de l'interface portant essentiellement sur le contenu, la présentation et les modalités des échanges d'informations entre l'utilisateur et le système. Après quelques définitions et la proposition d'une taxonomie qui permet de mettre en perspective les recherches actuelles et les avancées récentes sur les interfaces adaptatives, nous présentons quelques exemples représentatifs de ces travaux. Compte tenu du foisonnement des recherches sur la flexibilité des interfaces utilisateur et de l'ampleur du champ couvert par ce concept, nous avons limité la portée de cette étude aux interfaces utilisateur dont le comportement évolue au cours de l'interaction en fonction du profil de l'utilisateur et de ses évolutions tels qu'elles les « perçoivent ». Pour la même raison, l'adaptation de l'interaction au contexte d'utilisation n'est pas abordée ici. On trouvera dans (Bobillier-Chaumon et al., 2005) un panorama des recherches auxquelles a donné lieu la mise en œuvre du concept de flexibilité/adaptation en interaction homme-machine (IHM). Enfin, avant de conclure, nous présenterons brièvement nos travaux en cours sur l'adaptation à l'utilisateur.

**2. Définitions et proposition d'une taxonomie des formes d'adaptation**

Dans cette section, nous proposons une classification des différentes formes de mise en œuvre de la notion d'adaptation en interaction homme-machine, en précisant au fur et à mesure de leur apparition, la signification des principaux termes employés dans la suite de l'article. Cette classification est définie progressivement par raffinements successifs.

*2.1. Classification générale*

Nous distinguons avec A. Jameson (Jameson, 2003) deux types principaux d'adaptation, en fonction du but et de la finalité visés : adaptation au profil de l'utilisateur courant ou adaptation à l'environnement dans lequel se situe l'interaction. On utilise fréquemment le terme ***plasticité*** pour désigner cette seconde forme d'adaptation qui n'est pas évoquée dans la suite de l'article ; voir par exemple (Thévenin et al., 2003). La mise en œuvre du concept de plasticité vise essentiellement à résoudre les problèmes posés par la mobilité de l'utilisateur qui conduit celui-ci à utiliser des dispositifs d'interaction variés aux capacités d'échanges différentes de celles des stations de travail classiques (e.g., PDAs, ordinateurs, téléphones cellulaires, ordinateurs portables au sens de 'wearable', etc.). La plasticité mise en œuvre dans la majorité des solutions proposées porte esentiellement sur l'adaptation des modalités d'interaction aux ressources limitées offertes par les différents équipements et aux contraintes qu'impose la mobilité sur les capacités perceptives et motrices de l'utilisateur ; voir, par exemple, (Calvary et al., 2005).

L'adaptation à l'utilisateur courant, appelée aussi ***personnalisation*** de l'interface, peut s'effectuer de deux manières différentes. Si l'utilisateur a la possibilité d'adapter l'interaction à ses préférences personnelles, l'interface est dite configurable (*customizable*) ou ***adaptable***. Si le système est capable d'adapter son comportement aux besoins, capacités et préférences de l'utilisateur courant pendant l'interaction, grâce à ses capacités de perception et d'interprétation de l'interaction et de son contexte, l'interface est dite ***adaptative***.

Dans le premier cas, l'adaptation est généralement à l'initiative de l'utilisateur. Elle s'effectue le plus souvent sur la base d'un modèle de l'utilisateur prédéfini, donc statique (i.e., défini une fois pour toutes). Ces modèles statiques sont utilisés pour définir les options proposées à l'utilisateur. Ils sont généralement construits à partir d'une classification grossière des utilisateurs en un nombre limité de catégories ; ils ne peuvent donc rendre compte des caractéristiques individuelles spécifiques des utilisateurs ; voir dans (Shneiderman, 1987), le chapitre consacré à la classification des utilisateurs en trois classes principales (ou *stereotypes*).

Dans le second cas, l'adaptation est dynamique, et l'initiative des évolutions du comportement de l'interface utilisateur est souvent entièrement à la charge du système ; elle est alors totalement transparente pour l'utilisateur. Par exemple, les « Smart Menus », qui ont fait leur apparition dans les applications commercialisées destinées au grand public (voir Windows 2000), masquent les items rarement (ou jamais) sélectionnés par l'utilisateur courant qui n'a d'autre contrôle sur cette option de personnalisation que de la désactiver. Plus rarement, des possibilités de contrôler l'évolution de l'interface sont offertes à celui-ci, le partage des initiatives entre le système et lui variant dans des proportions importantes d'un logiciel à l'autre. Par exemple, les résultats d'un filtrage des dépêches de l'AFP (diffusées sur le Web) en fonction des intérêts de l'utilisateur, peuvent être présentés sous la forme d'une liste ordonnée des titres de la totalité des dépêches ; l'algorithme de filtrage peut aussi prendre l'initiative de supprimer certaines d'entre elles (e.g., en appliquant un critère quantitatif de redondance), sans en informer l'utilisateur (transparence).

Contrairement aux auteurs de (Mobasher et al., 2000), nous utilisons le terme 'personnalisation' uniquement pour désigner l'adaptation dynamique de l'interface au profil de l'utilisateur courant.

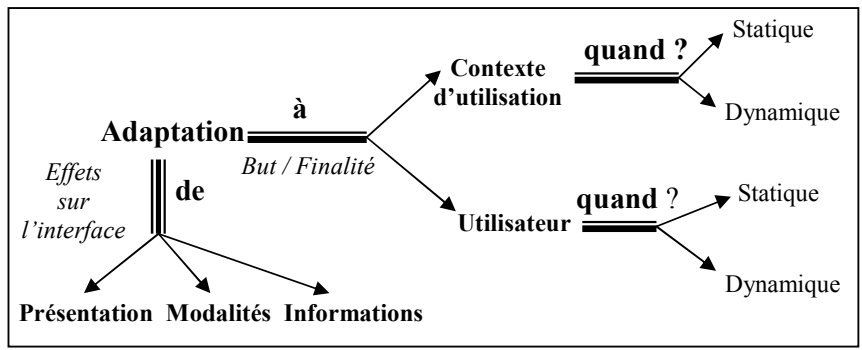

**Figure 1.** *Classification générale des formes de mise en œuvre du concept d'adaptation en IHM. Voir (Simonin, Carbonell, 2004).*

La figure 1 résume cette classification générale. On trouvera dans (Vanderdonckt et al., 2005) la présentation d'une taxonomie plus vaste qui couvre, outre l'adaptation statique ou dynamique au profil de l'utilisateur, celle à la situation d'interaction et au contexte d'utilisation. L'objectif des auteurs est de recenser, à l'intention des concepteurs d'interfaces adaptables, adaptatives ou plastiques, un panorama détaillé des choix de conception que permettent les avancées scientifiques et technologiques récentes.

A noter que l'hypothèse générale suivante, partagée par de nombreux chercheurs en IHM, contribue à l'essor actuel des recherches sur les interfaces adaptatives : doter les systèmes interactifs de capacités de dialogue analogues à celles que manifestent les humains en situation de communication pourrait permettre d'éliminer les problèmes d'utilisation que rencontrent les utilisateurs des logiciels grand public actuels, en particulier les novices, et d'obtenir des interfaces utilisateur vraiment transparentes. Or la faculté d'adaptation à l'interlocuteur est une composante de l'intelligence humaine qui joue un rôle primordial dans la communication face à face où l'on observe que le locuteur s'adapte constamment à ce qu'il perçoit des capacités, des connaissances et des intentions de son interlocuteur.

Cette hypothèse est discutable pour de nombreuses raisons. En particulier, il n'est pas certain que les utilisateurs souhaitent une interaction logicielle anthropomorphique (Amalberti et al., 1993). En outre, doter le système d'un comportement anthropomorphique va à l'encontre de l'une des recommandations de base de l'ergonomie des logiciels, dans la mesure où la mise en œuvre de cette stratégie peut induire l'utilisateur à supposer que le système possède l'ensemble des capacités humaines, dont l'intelligence, et donc favoriser des attentes et des exigences de la part de celui-ci qu'aucun système interactif n'est en mesure de satisfaire aujourd'hui ni même à moyen terme ; ce risque de l'anthropomorphisme en IHM a été signalé très tôt (Shneiderman, 1987). Cette discussion sera poursuivie dans la dernière partie de l'article où nous proposons quelques pistes de recherche.

D'autres critères permettent de raffiner cette classification grossière des interfaces utilisateur adaptatives. La perception qu'a le système de l'utilisateur et le modèle qui en résulte peut être centrée, selon l'application considérée, sur des caractéristiques individuelles de nature différente. Nous détaillons ces critères dans le paragraphe suivant.

## 2.2. Enrichissement de la classification générale

Les dimensions de la classification présentée dans la figure 1 ne rendent pas compte entièrement de la diversité des directions de recherche explorées jusqu'à présent concernant l'adaptation de l'interaction au profil de l'utilisateur. Elles doivent être complétées. Avant de donner, à titre d'illustration, des exemples de travaux représentatifs des différents courants de recherche actuels, nous précisons ici ces dimensions complémentaires :

– La nature des caractéristiques de l'utilisateur ciblées par l'adaptation.

– La répartition de l'initiative et du contrôle de l'adaptation entre celui-ci et le système.

– Le rôle de l'adaptation, c'est-à-dire la nature de l'activité de l'utilisateur qu'elle tente de faciliter et rendre plus efficace et donc, indirectement, le type d'application considéré.

*2.2.1. Diversité des constituants du profil et du modèle utilisateur*

Les dimensions de l'individualisation de l'interaction que permettent les modèles de l'utilisateur, statiques ou dynamiques, élaborés récemment sont nombreuses et d'une diversité remarquable. Les composantes suivantes de la personnalité de l'utilisateur courant ont fait l'objet de nombreuses modélisations :

– Ses *connaissances*, ses *compétences* et son *expérience*, ses *croyances*.

– Ses *intérêts*, ses *préférences*. Ces deux premières dimensions ont suscité les modélisations statiques et dynamiques les plus nombreuses.

– Ses *habitudes*, par exemple, les séquences d'actions qu'il effectue de façon routinière sur l'interface. Cette dimension ne peut donner lieu qu'à des modèles dynamiques puisqu'il s'agit d'une adaptation au comportement de l'utilisateur.

– Ses *capacités physiques*. Le handicap physique est un domaine de recherche à part entière qui privilégie la mise en œuvre du concept d'adaptabilité, compte tenu de la stabilité temporelle à court terme de la majorité des déficits perceptifs et moteurs. Une des réalisations les plus abouties est le prototype développé dans le cadre du projet européen AVANTI (Stephanidis, 2001). Cette dimension ne sera plus évoquée dans la suite de l'article puisqu'elle privilégie la conception d'interfaces utilisateur adaptables plutôt qu'adaptatives.

– Ses *intentions* et *buts* courants. Pour offrir à l'utilisateur une assistance efficace à la réalisation de ses intentions, il faut que le système soit capable de détecter de façon fiable les intentions de l'utilisateur sans que celui-ci ait besoin de les préciser au système, donc à partir des seules informations contextuelles disponibles. Modéliser dynamiquement les intentions de l'utilisateur courant représente un objectif ambitieux, extrêmement difficile à atteindre pour l'instant, sauf dans des contextes applicatifs particuliers ; ce qui peut expliquer pourquoi seules quelques recherches à caractère préliminaire et de portée limitée aient été menées dans cette direction jusqu'à présent. Voir par exemple (Conati et al., 1997) qui présente l'utilisation d'un réseau bayésien pour identifier et évaluer la stratégie de résolution de problème mise en œuvre par l'apprenant à partir des actions que celui-ci effectue sur l'interface du tuteur intelligent ANDES ; le domaine de connaissances considéré est la physique newtonienne.

– Ses *états psychologiques*, en particulier ses émotions (anxiété, stress, plaisir, etc.). Les recherches menées dans cette direction sont encore embryonnaires, mais elles sont susceptibles de se développer rapidement sous l'impulsion des travaux centrés sur l'apport à l'interaction d'agents conversationnels animés, ou ACAs, (voir, par exemple, Bevacqua, Pelachaud, 2004), et le soutien institutionnel apporté aux recherches sur l'interaction « affective » ; voir, notamment les objectifs du réseau européen d'excellence HUMANE qui compte plusieurs équipes possédant une solide expérience dans le domaine des ACAs (HUMANE, 2003).

Cette liste n'est pas exhaustive. On pourrait y inclure, par exemple, le rôle professionnel de l'utilisateur, son rôle social, etc. Mais ces dimensions représentent des caractéristiques individuelles relativement stables temporellement. Elles relèvent donc d'une adaptation de l'interaction statique plutôt que dynamique et, par conséquent, se situent en dehors du champ que cet article vise à couvrir.

A noter que le choix de la composante du profil utilisateur sur laquelle porte la personnalisation n'est pas indépendant du domaine d'application envisagé. Par exemple, les aides en ligne, les didacticiels et les Environnements Interactifs d'Apprentissage Assisté par Ordinateur, qui tentent d'adapter dynamiquement leur comportement au profil cognitif du novice ou de l'apprenant, se focalisent sur la modélisation des connaissances de l'utilisateur courant, de ses compétences spécifiques ou de son expérience personnelle. En revanche, les navigateurs Internet adaptatifs tentent le plus souvent de cerner les intérêts, goûts et préférences de chaque utilisateur.

*2.2.2. Répartition du contrôle de l'adaptation entre l'utilisateur et le système*

L'initiative et le contrôle de l'évolution de l'interface peuvent être du ressort exclusif de l'utilisateur. C'est le cas de la plupart des interfaces adaptables qui offrent à l'utilisateur un ensemble d'options et de possibilités de paramétrage sous différentes formes, par exemple : fichiers texte de configuration associés aux utilitaires des premiers systèmes d'exploitation (e.g., Unix), menus 'Affichage' et 'Fenêtre' des logiciels destinés au grand public (e.g., Word), ou item 'Préférences' de nombreux navigateurs (e.g., Netscape).

Le contrôle de l'évolution du comportement de l'interface peut également être assuré entièrement par le système. Ainsi, certains prototypes de recherche prennent-ils en charge la sélection du profil utilisateur au sein d'un ensemble plus ou moins riche de stéréotypes prédéfinis, en vue d'assurer la transparence de l'adaptation. Cette solution a été souvent adoptée, notamment, pour les applications destinées au grand public, telles que l'assistance à la recherche d'informations et la navigation sur Internet ou les logiciels éducatifs, car cette catégorie d'utilisateurs est censée rechercher en priorité et privilégier la transparence de l'interaction. Par exemple, c'est cette solution que les premiers prototypes de filtrage des informations sur Internet en fonction des intérêts de l'utilisateur courant ont mise en œuvre (Lieberman, Maulsby, 1996). Les environnements expérimentaux d'apprentissage adaptatifs des années 90 l'ont également adoptée pour la mise en œuvre de modèles complexes des connaissances de l'apprenant ; voir, par exemple, le modèle décrit dans (Paiva, Self, 1994 ; Paiva et al., 1994). A noter que la transparence de l'évolution du comportement du système a été également choisie pour les versions récentes de Pack Office ; voir les « Smart Menus » évoqués dans la section 2.1.

*2.2.3. Rôle de l'adaptation*

La finalité de l'adaptation dynamique au profil de l'utilisateur courant est généralement d'assister celui-ci dans la réalisation des objectifs et des activités qui motivent son interaction avec le logiciel adaptatif.

Un modèle centré sur les intérêts et les préférences de l'utilisateur permet de fournir à celui-ci une assistance explicite ou implicite à des activités de navigation et de recherche d'informations. Les systèmes « conseillers » (*recommending systems*), par exemple, assistent explicitement l'utilisateur dans ses prises de décisions en lui proposant des solutions et en argumentant leur bien-fondé (Ardissono, Goy, 2000 ; Fink, Kobsa, 2000). En revanche, les systèmes de filtrage des informations diffusées sur Internet, qui sélectionnent les informations présentées à l'utilisateur en fonction de ses intérêts, l'assistent implicitement dans sa recherche d'informations (Ardissono et al., 2001).

Un modèle de l'utilisateur capable d'évaluer les connaissances du novice ou de l'apprenant et de suivre leur évolution au cours de l'interaction permet de fournir aux utilisateurs d'un didacticiel, des informations pertinentes au bon moment et dans une présentation qui lui en facilitera la compréhension et l'assimilation (Mitchell et al., 1994 ; Brajnik, Tasso, 1994).

**3. Exemples de mise en œuvre du concept d'adaptation dynamique en IHM**

Nous décrivons dans cette section quelques exemples représentatifs de mise en œuvre du concept d'adaptation dynamique au profil de l'utilisateur, plutôt que de survoler l'ensemble des travaux publiés. Les recherches dans ce domaine ont souvent donné lieu au développement de systèmes adaptatifs plus ou moins aboutis, du démonstrateur ou du prototype de laboratoire au système opérationnel, testé dans des situations réelles d'utilisation prolongée. La présentation suit l'ordre de la classification proposée dans la section 2.

*3.1. Diversité des composantes du profil utilisateur visées par l'adaptation*

*3.1.1. Activités de routine et habitudes*

Les recherches dans ce domaine, relativement anciennes, se sont orientées dans deux directions principales.

L'une vise à accroître l'efficacité de l'interaction en déchargeant l'utilisateur de tâches routinières comme la gestion des messages (Maes, 1994) ou le suivi de planning/agenda (Mitchell et al., 1994). Généralement, l'utilisateur garde un contrôle sur la prise en charge de ces activités par le système. Par exemple, SwiftFile, logiciel téléchargeable sur le site de recherche d'IBM, assiste l'utilisateur dans le rangement des messages qu'il reçoit, en lui proposant, pour chaque nouveau message, trois dossiers qu'il considère comme les plus pertinents. La sélection de ces dossiers s'appuie sur une analyse statistique du contenu des messages et des pratiques de l'utilisateur (Segal, Kephart, 1999).

L'autre direction de recherche porte sur l'adaptation de la présentation des informations, principalement des menus, aux habitudes de travail de chaque utilisateur. Outre les « Smart Menus » (voir section 2.1), on peut citer les « Split Menus » proposés dans (Sears, Shneiderman, 1994) qui mettent en œuvre une stratégie différente. Cette stratégie consiste, lors de la sélection d'un menu, à afficher en début de liste (i.e., en haut de l'écran, sous l'en-tête du menu), les items les plus fréquemment sélectionnés par l'utilisateur, ce qui, d'après l'évaluation réalisée par les auteurs, accroît les performances (temps de sélection) et la satisfaction des utilisateurs ; comme pour les « Smart Menus », il s'agit d'une option de l'interface que l'utilisateur ne peut qu'activer ou désactiver.

*3.1.2. Capacités cognitives, connaissances, compétences et expérience*

L'adaptation de l'interaction aux connaissances et compétences de l'utilisateur courant a suscité de nombreux travaux de recherche. Elle représente l'une des mises en œuvre les plus étudiées en IHM du concept d'adaptation dynamique.

Le projet Lumiere (Horvitz et al., 1998) mérite une attention particulière en raison de l'influence que ses résultats ont exercée sur l'évolution des logiciels Microsoft destinés au grand public. Une version réduite (en termes de fonctionnalités) du prototype développé dans le cadre de ce projet a été intégrée aux aides en ligne des logiciels de Pack Office à la fin des années 90. Ces aides en ligne prennent l'initiative de proposer à l'utilisateur une ou plusieurs rubriques thématiques d'aide qu'elles jugent susceptibles de lui être utiles sur la base d'une analyse des actions récentes qu'il a effectuées sur le logiciel et de l'action en cours d'exécution ; pour accéder aux informations d'aide correspondant à ces rubriques, il suffit de sélectionner l'une d'entre elles.

Cette stratégie est discutable dans la mesure où les interventions « spontanées » de l'aide risquent d'interférer avec l'activité principale de l'utilisateur, à savoir l'interaction avec le logiciel, et donc d'être perçues par lui comme des interruptions intempestives, voire irritantes, de cette activité. En outre, anticiper à bon escient les besoins d'informations d'aide de l'utilisateur requiert des compétences qui sont aujourd'hui encore hors de la portée des systèmes dits « intelligents ». Enfin, au cas particulier, le comportement et les performances de la version implantée dans Pack Office sont trop limitées par rapport à celles du prototype issu du projet Lumiere pour que l'on puisse se fonder sur l'analyse des réactions des utilisateurs de Pack Office pour déterminer l'utilité et l'utilisabilité (*usability*[1]) d'une aide en ligne qui tente d'anticiper les besoins en informations d'aide de l'utilisateur novice.

On trouvera dans la section 3.2.3 un exemple plus classique, représentatif des environnements adaptatifs d'enseignement et des tuteurs « intelligents » qui mettent à profit un suivi de l'évolution des connaissances de l'apprenant au cours de l'interaction pour améliorer l'efficacité pédagogique du système.

---

[1] Ce terme introduit par J. Nielsen fait référence à un ensemble de critères qui permettent d'évaluer l'effort nécessaire pour utiliser un logiciel de façon satisfaisante (ISO 9126, 1991).

*3.1.3. Etats émotionnels*

Dans le contexte des travaux sur l'interaction homme-machine affective, l'objectif qui sous-tend le plus souvent la modélisation de l'état émotionnel courant de l'utilisateur est moins d'adapter le comportement du système à cet état et à ses fluctuations au cours de l'interaction que d'agir sur lui s'il nuit à la réalisation des buts de l'utilisateur.

Les travaux de recherche présentés dans (Schäfer, Weyrath, 1997) s'inscrivent dans une perspective d'adaptation dynamique du comportement d'un système d'assistance à l'état émotionnel de l'utilisateur, en particulier aux effets de cet état sur ses capacités cognitives. READY (Resource-Adaptive Dialog System) est un prototype qui tente d'évaluer le degré de stress et d'anxiété des auteurs d'appels à un service téléphonique d'assistance aux urgences (incendies) afin, d'une part, d'adapter le contenu et la présentation des messages oraux à l'altération temporaire des capacités cognitives des appelants (i.e., mémorisation, élocution et maîtrise de la situation) due aux effets négatifs de ces émotions et d'une intoxication éventuelle et, d'autre part, de tenir compte dans la reconnaissance de leurs énoncés de la dégradation de leur élocution. La détection de l'état émotionnel de l'appelant utilise un réseau bayésien dynamique pour analyser son rythme respiratoire, son élocution et les bruits de fond. Le comportement du système met à profit les connaissances issues de l'étude d'un corpus de dialogues enregistrés dans un centre d'appels téléphoniques des pompiers.

Des travaux plus récents visent non pas à adapter le comportement du système à l'état émotionnel de l'utilisateur courant mais à agir sur cet état. Ainsi, (Prendinger et al., 2005) décrit une étude où le comportement empathique d'un agent conversationnel animé réduit significativement l'anxiété de l'utilisateur pendant un entretien d'embauche simulé. L'objectif est d'entraîner celui-ci à contrôler ses émotions négatives au cours de ce type d'entretien par un apprentissage (conditionnement) de type behaviouriste. La détection de l'état émotionnel (*arousal*) de l'utilisateur est effectuée au moyen de capteurs physiologiques du rythme cardiaque (électromyographie) et de la conductance de la peau (capteur galvanique).

*3.2. Répartition de l'initiative et du contrôle de l'adaptation – Rôle de l'adaptation*

Nous avons rassemblé dans une même section ces deux dimensions car elles sont liées au domaine d'application envisagé. Les recherches sur l'adaptation dynamique à l'utilisateur privilégient les classes d'applications grand public suivantes qui se prêtent naturellement à une personnalisation de l'interaction :

– Les systèmes et agents « conseillers » (*recommending systems*), destinés souvent à des sites Web commerciaux.

– Les systèmes d'assistance personnalisée à la recherche d'informations sur Internet.

– Les systèmes à vocation didactique : environnements d'enseignement à distance, didacticiels, tutoriels et aides en ligne.

Les exemples présentés dans la suite de cette section ont été regroupés en fonction du domaine applicatif ciblé.

*3.2.1. Les systèmes conseillers*

En raison de leur vocation, les systèmes conseillers détiennent l'initiative et le contrôle de l'adaptation puisque, par définition, leur rôle est d'assister l'utilisateur dans un choix, une prise de décision. Ils se distinguent les uns des autres par la nature des informations qu'ils utilisent pour modéliser les intérêts et préférences de l'utilisateur et par la méthode qu'ils appliquent pour acquérir ces informations.

Par exemple, MovieCentral, site de Entertainment Inc. qui fournit des recommandations personnalisées sur les films à l'affiche, initialise le modèle de l'utilisateur courant U à partir de données explicites, à savoir l'évaluation par celui-ci de dix films (à l'aide de notes comprises entre 1 et 10). Le système, qui applique une approche dite « collaborative », identifie, dans sa base des profils utilisateur, le groupe d'utilisateurs dont les vecteurs de notes sont les plus proches de celui de U. Ce qui permet au système de prédire, en réponse à une requête de U sur un film donné, si ce film lui plaira ou non.

Certains sites imposent à U d'expliciter ses critères d'appréciation plutôt que d'évaluer un ensemble limité de produits. D'autres encore construisent le profil de l'utilisateur de façon incrémentale. C'est le cas du prototype présenté dans (Burke, 2001). Le système S propose à U un produit, un restaurant au cas particulier, et invite celui-ci à critiquer et évaluer qualitativement cette proposition initiale. S fait ensuite une nouvelle proposition qui tient compte des critiques de U, et répète cette procédure jusqu'à ce que U ne formule plus de critiques sur la proposition courante. Cette démarche itérative permet à l'utilisateur de raffiner progressivement et de façon naturelle ses critères de choix, et au système d'obtenir au terme du raffinement un profil utilisateur pertinent et précis sans imposer à celui-ci un questionnaire initial artificiel et fastidieux.

*3.2.2. L'assistance à la navigation et à la recherche d'informations sur Internet*

Le rôle principal des systèmes adaptatifs qui visent ce domaine d'application consiste à filtrer la masse d'informations disponibles sur la Toile en fonction des intérêts de l'utilisateur courant. Si le système de personnalisation ne présente à l'utilisateur que les informations qui sont censées l'intéresser, l'adaptation est transparente pour celui-ci, il ne peut donc la contrôler. En revanche, si le système se borne à ordonner les résultats de la recherche en fonction de ce qu'il suppose être les intérêts de l'utilisateur, celui-ci peut contrôler l'adaptation et, par ses réactions, contribuer à améliorer la pertinence du modèle sur lequel s'appuie la personnalisation.

Parmi les travaux de recherche publiés sur le filtrage adaptatif d'informations, l'une des études les plus intéressantes est celle décrite dans (Billsus et Pazzani, 2000). C'est une recherche parmi les plus abouties puisqu'elle a donné lieu à un prototype qui a fait l'objet d'une évaluation à grande échelle en situation d'utilisation réelle (recueil des traces de 2000 utilisateurs). C'est aussi l'une des plus évoluées puisqu'elle met en œuvre un modèle à court terme et un modèle à long terme des intérêts de l'utilisateur, et utilise un feedback explicite et implicite fourni par celui-ci pour la mise à jour de ces deux modèles. Le problème de la plasticité est également abordé puisque deux versions du prototype ont été réalisées, l'une pour le Web, l'autre pour PDAs et téléphones portables.

Les auteurs ont réalisé la conception, le déploiement et l'évaluation d'un système interactif S d'accès aux actualités (i.e., les *news*). La version Web de S fournit, en réponse à une requête de l'utilisateur U, exprimée sous la forme d'une expression en langue naturelle ou d'une suite de mots-clés, la liste des titres des bulletins d'information susceptibles de satisfaire la requête, ordonnés en fonction du profil de U ; S affiche également, pour chaque titre, son score de prédiction de l'intérêt de U. La sélection d'un titre dans cette liste entraîne l'affichage du bulletin d'information correspondant. U peut alors, s'il le souhaite, fournir un jugement explicite sur l'information affichée en sélectionnant l'un des items suivants : « Intéressant », « Inintéressant », « Encore », « Déjà connu », « Demande d'explication ». S'il sélectionne le dernier item, S répond en affichant : son score de prédiction pour cette information accompagné d'une phrase d'explication en langue naturelle, le nombre d'autres utilisateurs qui l'ont évaluée et leur intérêt pour l'information considérée.

L'algorithme d'apprentissage qui construit le modèle global d'un utilisateur et le met à jour utilise les jugements explicites de U (« Intéressant », « Inintéressant »), les titres non sélectionnés par lui, les bulletins qu'il a lus (temps passé). Pour prédire l'intérêt qu'un nouveau document est susceptible de provoquer de la part de U, l'algorithme utilise une mesure de similitude[2] entre ce document et ceux qui ont suscité l'intérêt de U. A partir de ces informations, S construit deux modèles complémentaires de U : un modèle à court terme qui tient compte de l'intérêt de U pour les sujets et événements récents de l'actualité, et un modèle à long terme centré sur les préférences générales de U indépendantes de ses intérêts conjoncturels. Le calcul du score à court terme d'un nouveau document utilise l'algorithme « du plus proche voisin », et l'évaluation de son score à long terme un classificateur bayésien. Ces deux scores sont combinés dans un algorithme global multi-stratégies qui définit, sous la forme d'un score, la prédiction de S concernant l'intérêt global de U pour le document considéré.

Les principales avancées récentes portent sur le filtrage collaboratif, c'est-à-dire l'utilisation des profils d'autres utilisateurs pour compléter et préciser le profil d'un nouvel utilisateur (voir, par exemple, Kleinberg, Sandler, 2004). Le profil initial du nouvel utilisateur peut être établi, par exemple, à partir des réponses à un questionnaire initial (Ardissono et al., 2001).

---

[2] Les bulletins d'information sont convertis en vecteurs *tf-idf*, et la similitude entre deux bulletins est calculée par comparaison de leurs vecteurs respectifs à l'aide de la mesure définie dans (Salton, 1989).
Salton, G. (1989). *Automatic Text Processing*. Addison-Wesley.

*3.2.3. Environnements d'enseignement*

Les environnements d'assistance à l'enseignement à distance suscitent depuis plusieurs décennies une activité de recherche intense et constituent un domaine à part entière qui accorde une place importante aux travaux sur la modélisation de l'apprenant et la personnalisation des apports de connaissances, des exercices et de l'évaluation. On observe, en particulier, une grande diversité dans :

– les stratégies de contrôle de l'apprentissage, qui varient d'un guidage strict du parcours de l'apprenant à une liberté plus ou moins grande d'exploration des unités.

– l'aspect de l'interaction sur lequel porte l'adaptation dynamique, navigation dans les unités d'enseignement, choix des connaissances et concepts nouveaux présentés, et modulation de leur présentation, du contenu des suggestions et retours système, ou encore de la stratégie d'intervention du système « tuteur ».

– les caractéristiques de l'apprenant sur lesquelles portent la modélisation et l'adaptation : connaissance des contenus de l'enseignement, style d'apprentissage, motivation, activité en cours.

Le prototype ELM-ART (Weber, Specht, 1997), bien que relativement ancien, est intéressant ici, car il offre à l'apprenant un guidage flexible pour l'apprentissage de LISP. Ce guidage tient compte des compétences de celui-ci et des pré-requis des unités de cours. Il propose à l'apprenant deux formes de recommandations. L'une est une liste d'unités associées chacune à une icône dont la couleur exprime la force de la recommandation ; l'apprenant dispose donc d'une relative liberté d'exploration. L'autre recommande une seule unité ; elle constitue un guidage strict de l'apprenant.

**4. Notre contribution**

Nos travaux actuels portent sur les aspects ergonomiques, encore peu étudiés, de l'adaptation dynamique à l'utilisateur. L'objectif de l'étude expérimentale en cours de réalisation est d'évaluer l'apport effectif d'un modèle dynamique des connaissances et compétences de l'utilisateur à l'efficacité de l'interaction d'assistance. Le domaine d'application choisi est l'aide en ligne à l'utilisation des logiciels grand public, logiciels dont l'utilisation se maîtrise rapidement. Il est possible, dans ce contexte, d'observer une évolution importante des savoir-faire des novices sur une période de quelques dizaines de minutes, ce qui permet d'utiliser des protocoles expérimentaux classiques en IHM qui se bornent à étudier les comportements et réactions d'utilisateurs pendant une seule session d'une durée limitée.

Nous avons choisi un logiciel de création d'animations, Flash, peu connu des étudiants de Licence parmi lesquels sont sélectionnés les sujets. L'aide en ligne comprend plusieurs centaines de messages multimodaux (texte + graphique). Chaque message d'aide existe en trois versions : l'une, envoyée en réponse à la première requête du sujet, les deux autres en réponse à des requêtes ultérieures éventuelles de sa part pendant la session. L'une des deux dernières, plus détaillée, est destinée aux sujets ayant manifesté des difficultés à mettre en œuvre la version initiale du message, l'autre, réduite à un bref rappel procédural, aux sujets qui font preuve d'une progression rapide dans la maîtrise de Flash et ont mis en œuvre correctement les indications figurant dans la première version du message qui lui a été envoyée.

La simulation de l'interface adaptative du système d'aide fait appel à la technique dite du magicien d'Oz : c'est le compère qui choisit la version du message qui sera envoyée au sujet en réponse à une requête de sa part. Le rôle du compère se limite à ce choix. L'envoi du message au sujet, la transmission au compère de la trace des interactions du sujet avec l'aide et avec Flash (sous forme de copies d'écran), ainsi que la sauvegarde de la trace de ces interactions, sont assurés par une plate-forme logicielle générique que nous avons développée pour mettre en œuvre la technique du magicien d'Oz. Ainsi, seules l'élaboration du modèle de l'utilisateur et sa mise en œuvre sont-elles confiées au compère.

La réalisation de cette étude s'appuie sur l'expérience que nous avons acquise dans le domaine de l'aide en ligne (Capobianco, Carbonell, 2003) et dans la mise en œuvre de la technique du magicien d'Oz (Carbonell, 2003).

## 5. Conclusions

Cet aperçu des recherches sur la mise en œuvre en IHM du concept d'adaptation dynamique à l'utilisateur montre d'abord la diversité des données à partir desquelles sont construits les modèles de l'utilisateur : questionnaires initiaux, trace des interactions, profils d'autres utilisateurs. Les techniques de modélisation sont également très variées : inférence logique (logiques modales, épistémiques, non monotones) ou inférence fondée sur des méthodes statistiques ou probabilistes (réseaux connexionnistes ou bayésiens, notamment).

On observe par ailleurs un déséquilibre entre, d'une part, le volume important des travaux consacrés à l'élaboration de modèles pertinents de l'utilisateur et à la mise en œuvre efficace du processus d'adaptation dynamique à l'utilisateur et, d'autre part, la rareté des études ergonomiques sur l'efficacité, la facilité d'utilisation et l'acceptation par les utilisateurs potentiels de cette nouvelle forme d'interaction qu'offrent les interfaces adaptatives.

Ce déséquilibre s'explique peut-être par les difficultés que présente l'étude ergonomique des systèmes d'interaction adaptatifs.

Ces difficultés sont essentiellement d'ordre méthodologique. Comment évaluer l'utilité et l'utilisabilité d'un système adaptatif qui nécessite un usage prolongé pour que l'utilisateur puisse se représenter les capacités exactes du système et que les effets d'un tel système sur son comportement et ses performances soient perceptibles ? La technique du magicien d'Oz est inapplicable dans un tel contexte. Il faut pouvoir disposer d'un prototype. Mais alors on risque d'évaluer les caractéristiques spécifiques d'une implémentation particulière de l'adaptation dynamique à l'utilisateur, plutôt que la mise en œuvre du concept d'interaction adaptative. En outre, quelle durée choisir compte tenu de la variabilité intra-individuelle élevée des comportements cognitifs et des jugements subjectifs ? Combien d'utilisateurs potentiels impliquer dans les études empiriques ou expérimentales pour obtenir des conclusions valides étant donné la variabilité inter-individuelle des capacités et processus cognitifs ? Comment évaluer la pertinence du modèle de l'utilisateur ? Quelles méthodes mettre en œuvre ? Quelles mesures utiliser ? (Chin, 2001) fournit quelques indications méthodologiques utiles pour effectuer l'évaluation empirique des modèles de l'utilisateur et des systèmes adaptatifs ; voir également (Herlocker et al., 2004).

Les interfaces adaptatives posent de nombreuses questions de recherche nouvelles quant à leur apport effectif à l'interaction. Quel contrôle donner à l'utilisateur sur l'évolution du comportement du système ? La transparence de l'adaptation est-elle souhaitable? Ne vaut-il pas mieux notifier l'utilisateur de l'évolution spontanée du comportement du système au cours de l'interaction et, si oui, sous quelle forme ? Les réponses à ces questions dépendent, en particulier, de l'activité de l'utilisateur à laquelle le système fournit une assistance, donc du domaine d'application ; elles dépendent également de la nature des caractéristiques individuelles modélisées, de la composante du comportement du système objet de l'adaptation. Il est impossible de répondre globalement à ces questions qui doivent être abordées dans le cadre d'un vaste programme de recherche.

## 12. Bibliographie